\documentclass[pra,twocolumn,superscriptaddress,showpacs]{revtex4}

\usepackage{graphics}           
\usepackage{bm}                 
\usepackage{amsmath}            

\begin{document}

\title[Thermal concurrence mixing in a 1D Ising model]{Thermal concurrence mixing in a 1D Ising model}

\author{D. Gunlycke}
\affiliation{Optics Section, Blackett Laboratory, Imperial College, London, SW7 2BW, UK.}

\author{S. Bose}
\affiliation{Centre for Quantum Computation, Clarendon Laboratory,
  Parks Road, Oxford OX1 3PU, UK.}

\author{V. M. Kendon}
\affiliation{Optics Section, Blackett Laboratory, Imperial College, London, SW7 2BW, UK.}

\author{V. Vedral}
\affiliation{Optics Section, Blackett Laboratory, Imperial College, London, SW7 2BW, UK.}

\date{\today}

\begin{abstract}
We investigate the entanglement arising naturally in a 1D Ising
chain in a magnetic field in an arbitrary direction. We find that
for different temperatures, different orientations of the magnetic
field give maximum entanglement. In the high temperature limit,
this optimal orientation corresponds to the magnetic field being
perpendicular to the Ising orientation ($z$ direction). In the low temperature limit, we
find that varying the angle of the magnetic field very slightly
from the $z$ direction leads to a rapid rise in entanglement. We
also find that the orientation of the magnetic field for maximum
entanglement varies with the field amplitude. Furthermore, we have
derived a simple rule for the mixing of concurrences (a measure of
entanglement) due to mixing of pure states satisfying certain conditions.
\end{abstract}


\maketitle

\section{Introduction}

Entanglement, the rather counterintuitive non-local correlations
exhibited by quantum systems, has recently become one of the
most valuable resources in quantum information processing
\cite{bennett00}. Over the past few years it has developed into a
quantifiable physical resource \cite{vedral97,bennett96,wootters98} in
an analogous manner to energy. Thus, the amount of entanglement present
naturally in complex physical systems (systems with many
interacting components) now becomes a relevant question to ask.
Condensed matter physicists have long investigated
correlations between parts of composite systems. Entanglement is the
``quantum'' or ``nonlocal'' part of these
correlations. As such, it can behave very differently from the
total correlations. For example, while correlations are averaged
on mixing states, entanglement generally decreases. The topic of
variation of entanglement in condensed matter systems with
respect to the variation of external parameters such as
temperature, field components etc., is a relatively
unexplored and potentially rich area of study. In this context,
as a simple initial model, Arnesen, Bose and Vedral have studied
the variation of entanglement with temperature and magnetic field
in a 1D isotropic finite Heisenberg chain \cite{arnesen00}. Prior
to that, Nielsen had investigated the entanglement between two
qubits interacting via the Heisenberg interaction at a nonzero
temperature \cite{nielsen98} and O'Connor and Wootters have
investigated the entanglement in the ground state of an
antiferromagnetic isotropic Heisenberg ring \cite{wootters00}. In
Ref. \cite{arnesen00}, the entanglement at a nonzero temperature,
being that of a thermal state, was called {\em thermal
entanglement}. Subsequently, Wang has studied the quantum
Heisenberg XY-model \cite{wang01} and the two-qubit anisotropic
XXZ-model \cite{wang01b} in a similar context.

In this paper, we are going to study the thermal entanglement in the
1D Ising model in an external magnetic field \cite{sachdev}. 
Ising-like interactions form the basic coupling in many proposals for
experimental systems that can be used to perform quantum computation,
see for example, \cite{cory96,kane98,mooij99,briegel00}.
The 1D Ising model describes a set of linearly arranged spins (qubits), each
interacting with its nearest neighbors by a coupling which is proportional to
$\sigma_z\otimes\sigma_z$. This coupling can be diagonalized in a
basis of disentangled states. Na\"{\i}vely one might think that
this implies a complete absence of entanglement in the Ising
model. However, an external magnetic field with a component,
however small, along a direction perpendicular to the $z$ axis is
sufficient to make the eigenstates entangled.

We start in section \ref{2-qubits} by considering analytically
the case of two qubits interacting via the Ising interaction in a
magnetic field orthogonal to the $z$ direction. In section
\ref{mixing}, we formulate a
theorem for the concurrence mixing due to occupation of both the
ground and the excited states. Next, in section
\ref{theta},  we consider numerically the variation of
entanglement with the orientation of magnetic field. Before
concluding, in section \ref{many} we show that the kinds of behavior
found for two qubits also hold for many qubits.

\section{Two qubits with Ising interaction}
\label{2-qubits}

The Hamiltonian for an isotropic quantum Ising model with nearest neighbor
couplings in an external magnetic field can in the most general
form be expressed as
\begin{equation}
  \label{eq:general}
  \hat{H} = J\sum_{<i,j>}\hat{\sigma}_z^i \hat{\sigma}_z^j +
  \vec{B}\cdot\sum_k\vec{\hat{\sigma}^k},
\end{equation}
where the indices $i$, $j$ and $k$ label the $N$ spins. Here we will
consider systems in one spatial dimension with periodic boundary
conditions so the $N$th spin also couples to the first spin. Thus we
have a qubit ``ring''. First we will consider the case of $N=2$. Our
Hamiltonian can then be written as
\begin{equation}
  \label{eq:Hamiltonian2}
  \hat{H} = 2J\,\hat{\sigma}_z\otimes\hat{\sigma}_z + \vec{B}\cdot(\vec{\hat{\sigma}}\otimes\hat{I}+\hat{I}\otimes\vec{\hat{\sigma}}).
\end{equation}
The usual form of the Ising model has  a magnetic field only along
the $z$ axis. This case has no entanglement at all, since the
Hamiltonian is diagonal in the standard disentangled basis
$\{|00\rangle,|01\rangle,|10\rangle,|11\rangle\}$, where
$|0\rangle$ stands for spin up and $|1\rangle$ stands for spin
down. However, in this paper we will consider the cases when the
magnetic field is not parallel to the $z$ axis.

\subsection{Orthogonal fields}

Let us first study the special case when the magnetic field is
perpendicular to the $z$ axis, say $\vec{B} = B\,\vec{x}$. Our
system is now described by the following Hamiltonian:
\begin{equation}
  \label{eq:Hamiltonian2Orthogonal}
  \hat{H} = 2J\,\hat{\sigma}_z\otimes\hat{\sigma}_z + B\,(\hat{\sigma}_x\otimes\hat{I}+\hat{I}\otimes\hat{\sigma}_x).
\end{equation}
We are going to investigate the entanglement in this two-qubit Ising
ring. In this paper, we will use the
squared concurrence \cite{wootters98,wootters00a}, called the tangle
$\tau$, as a measure of entanglement. To calculate this, first we need to define the product matrix $R$
of the density matrix, $\rho$, and its spin-flipped matrix, $\tilde{\rho} =
(\hat{\sigma}_y\otimes\hat{\sigma}_y)\rho^*(\hat{\sigma}_y\otimes\hat{\sigma}_y)$. Hence, we have
\begin{equation}
  \label{eq:productmatrix}
  R\equiv\rho\tilde{\rho} =\rho(\hat{\sigma}_y\otimes\hat{\sigma}_y)\rho^*(\hat{\sigma}_y\otimes\hat{\sigma}_y).
\end{equation}
Now concurrence is defined by
\begin{equation}
  \label{eq:concurrence}
  C = max\{\lambda_1-\lambda_2-\lambda_3-\lambda_4,0\},
\end{equation}
where the $\lambda_i$ are the square roots of the eigenvalues of $R$, in
decreasing order. In this method the standard
basis, $\{|00\rangle,|01\rangle,|10\rangle,|11\rangle\}$, must be
used. As usual for entanglement measures,
the tangle ranges from $0$ (no entanglement) to $1$, when the two
qubits are maximally entangled.

\begin{figure}
    \begin{minipage}{\columnwidth}
    \begin{center}
        \resizebox{1.0\columnwidth}{!}{\includegraphics{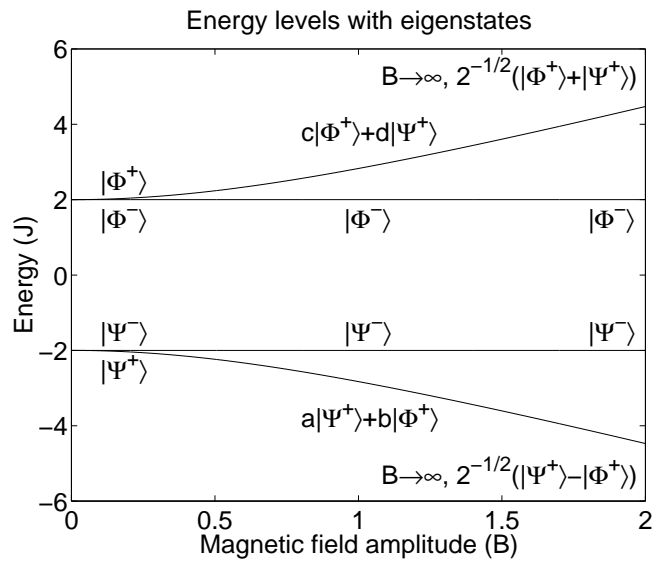}}
    \end{center}
    \end{minipage}
    \caption{Energy levels with corresponding eigenstates, a,b,c and d
      are functions of B, and the states $|\Phi\pm\rangle = (|00\rangle\pm|11\rangle)/\sqrt{2}$ and
      $|\Psi\pm\rangle = (|01\rangle\pm|10\rangle)/\sqrt{2}$ are the
      four Bell states.}
    \label{fig:2}
\end{figure}

 For finite temperatures we need the density matrix,
$\rho$, for a system which is at thermal equilibrium. This is
given by $\rho = e^{-\hat{H}/T}/Z$, where $Z = tr(e^{-\hat{H}/T})$ is the
partition function (using units where the Boltzmann
constant, $k_B = 1$). We then solve the time independent
Schr\"{o}dinger equation
for our qubits. The energy levels of our
Hamiltonian (\ref{eq:Hamiltonian2Orthogonal}) are, in rising
order, $-2\sqrt{J^2+B^2},-2J,2J,2\sqrt{J^2+B^2}$, as in
Fig. \ref{fig:2}.

\begin{figure}
    \begin{minipage}{\columnwidth}
    \begin{center}
        \resizebox{1.0\columnwidth}{!}{\includegraphics{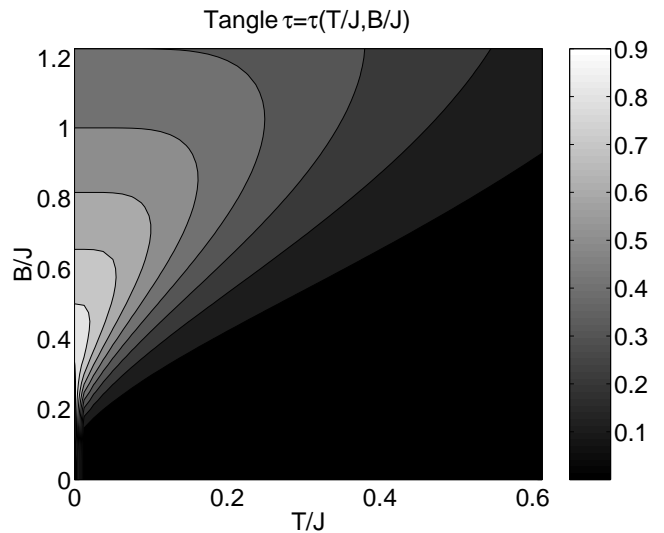}}
    \end{center}
    \end{minipage}
    \caption{Contour plot of the tangle of two qubits obeying an
      Ising Hamiltonian with coupling $J$, in a perpendicular
      magnetic field $B$, for temperatures $T$.}
    \label{fig:1}
\end{figure}

For zero temperature only the lowest energy level is
populated. The tangle of this pure state can easily be calculated
from the density matrix, for $B>0$,
\begin{equation}
  \label{eq:tangle}
  \tau = \frac{J^2}{J^2+B^2} = \frac{1}{1+\left(\frac{B}{J}\right)^2}.
\end{equation}
In Fig. \ref{fig:1}, a
contour plot of the tangle $\tau$, as a function of magnetic
field amplitude, $B$, and the temperature, $T$ is shown.

From Eq. (\ref{eq:tangle}), it is clear that the entanglement is
highest for nearly vanishing magnetic fields and decreases
with increasing field amplitude. However,
Eq. (\ref{eq:tangle}) is not valid for strictly $B=0$, in which
limit it seems to predict maximal entanglement. At precisely
$B=0$, in fact, no entanglement is present (the eigenstates are
same as those of the usual Ising Hamiltonian without any magnetic
field). Hence there is a {\em quantum phase transition}
\cite{sachdev} at the point B=0 when the entanglement jumps from
zero to maximal even for an infinitesimal increase of $B$. As we
will see later this point is only one point on a
transition line for $B$-fields along the $z$ axis.

  Let us now turn our attention to the more realistic case of non-zero
temperatures. For a general pure state only one of the
eigenvalues of Eq. (\ref{eq:productmatrix}) is non-zero and
therefore equal to the tangle. This statement is proved in
\emph{lemma 1} in the next section. For low temperature and
magnetic field, i.e. $B,T\ll J$, it is a good
approximation to assume that only the two lowest energy levels
are populated. This becomes clear when we look at Fig. \ref{fig:2}
in the regime $B\ll J$. The lowest two levels are much closer to
each other compared with their separation from the third lowest
energy level (i.e. the second excited state). Thus when the temperature
is low, only the lowest two levels appear in the state of the
system. We will find (\emph{theorem 1}, next section) that, in our case, the combination of the two lowest states also
combines their concurrences in the following weighted subtraction:
\begin{equation}
  \label{eq:weightconcurrence2}
  C = \max\{|w_0C_0 - w_1C_1|,0\},
\end{equation}
where the index $0$ refers to the ground state, while $1$ refers
to the excited state and $w_0$ and $w_1$ are the weights of the
ground and excited states respectively. The weights can be any
weights from the statistics, for example Maxwell-Boltzmann
statistics or Fermi-Dirac statistics. We call this {\em
concurrence mixing}. In our case, the first excited state is the Bell state,
 $|\Psi^{-}\rangle = (|01\rangle-|10\rangle)/\sqrt{2}$, which has
 $\tau = 1$, and Eq. (\ref{eq:weightconcurrence2}) reduces
to
\begin{equation}
\label{eq:reducedweightconcurrence2}
 C =|w_0\frac{J}{\sqrt{J^2+B^2}} - w_1|.
\end{equation}
In general, the first term in the above equation is larger than
the second, and in this case the concurrence decreases with
temperature as $w_0$ decreases and $w_1$ increases ({\em cf}
Fig. \ref{fig:1}). In Fig. \ref{fig:1} we also see that, for a given
temperature, the entanglement can be increased by adjusting the
magnetic field and is generally largest for some intermediate
value of the magnetic field. This effect can be understood by
noting that $w_0$ increases with increasing $B$ as the
energy separation between the levels increase, but
$J/\sqrt{J^2+B^2}$ decreases. As a result the combined function
reaches a peak as we vary $B$ and decreases subsequently,
inducing analogous behavior for the concurrence.

\section{Concurrence mixing}
\label{mixing}

In this section, we are going to formulate and prove a useful
concurrence mixing theorem. We begin with a lemma which illustrates
the method used in the theorem. The results of \emph{lemma 1} appear in Ref. \cite{wootters98}.
\\
\\
\textbf{Lemma 1:} \emph{Let $\rho$ be a pure density matrix. Then
the product matrix $R = \rho\tilde{\rho}$, where $\tilde{\rho} =
(\hat{\sigma}_y\otimes\hat{\sigma}_y)\rho^*(\hat{\sigma}_y\otimes\hat{\sigma}_y)$, has
only one non-zero eigenvalue, and its value is the concurrence
squared, i.e. the tangle. For a general pure state,
$|\alpha\rangle = a|00\rangle + b|01\rangle + c|10\rangle +
d|11\rangle$ the concurrence is $C = 2|ad-bc|$ or, written as a
Schmidt decomposition, $C = 2|c_0c_1|$, where $c_i$ are the two
Schmidt coefficients.}
\\
\\
\textsf{Proof:} Consider a general pure density matrix $\rho =
|\alpha\rangle\langle\alpha|$. By writing out the product matrix
\begin{equation}
  \label{eq:Lem1}
  \rho\tilde{\rho} =
|\alpha\rangle\langle\alpha|\hat{\sigma}_y\otimes\hat{\sigma}_y|\alpha\rangle\langle\alpha|\hat{\sigma}_y\otimes\hat{\sigma}_y,
\end{equation}
we see directly that $|u_0\rangle = |\alpha\rangle$ is an
eigenstate with eigenvalue
$|\langle\alpha|\hat{\sigma}_y\otimes\hat{\sigma}_y|\alpha\rangle|^2$. It is
always possible to find three more vectors
$|\alpha_k^\perp\rangle, k = 1,2,3$ all of them linearly independent of
each other and orthogonal to $|\alpha\rangle$. Thus the remaining
three eigenvectors can be written as $|u_k\rangle =
\hat{\sigma}_y\otimes\hat{\sigma}_y|\alpha_k^\perp\rangle, k = 1,2,3$, all
with eigenvalue zero. From Eq. (\ref{eq:concurrence}) we now get
the concurrence to be $C =
|\langle\alpha|\hat{\sigma}_y\otimes\hat{\sigma}_y|\alpha\rangle| =
2|ad-bc|$. Thus, for a pure state, concurrence can be defined as
the absolute expectation value of the operator
$\hat{\sigma}_y\otimes\hat{\sigma}_y$. By
tracing out one qubit and solving for the eigenvalues, which are
equal to the square of the Schmidt coefficients, $c_i$, of the
remaining density matrix, we find that the concurrence also can
be written as $C = 2|c_0c_1|$.
\\
\\
\textbf{Theorem 1:} \emph{Consider two pure states of the same system
$|\alpha_m\rangle$ and $|\alpha_n\rangle$. If the spin-flip
overlap is zero, i.e.}
\begin{equation}
  \label{eq:condition}
  \langle\alpha_m|\hat{\sigma}_y\otimes\hat{\sigma}_y|\alpha_n\rangle = 0,
\end{equation}
\emph{then the concurrence of the mixture of the two pure states,
  with weights $w_i$, can be expressed as}
\begin{equation}
  \label{eq:theorem}
  C_{mixed} = |w_mC_m - w_nC_n|.
\end{equation}
\\
\\
\textsf{Proof:} Let $\rho_i = |\alpha_i\rangle\langle\alpha_i|$, $i = m,n$, be our two pure states.
From our lemma, we have
\begin{eqnarray}
  \label{eq:1}
  \rho_i\tilde{\rho_i}|u_{i0}\rangle &&= C_i^2|u_{i0}\rangle
  \\
  \label{eq:2}
  \rho_i\tilde{\rho_i}|u_{ik}\rangle &&= 0,\hspace{1.5cm} k = 1,2,3
\end{eqnarray}
where $|u_{i0}\rangle = |\alpha_i\rangle$ and $|u_{ik}\rangle =
\hat{\sigma}_y\otimes\hat{\sigma}_y|\alpha_{ik}^\perp\rangle$. Let us write
our mixed state, $\rho$, as a weighted average of the pure density
matrices, $\rho = w_m\rho_m + w_n\rho_n$. Since $\tilde{\rho}$ is
only a linear transformation of $\rho$, we also have
$\tilde{\rho} = w_m\tilde{\rho_m} + w_n\tilde{\rho_n}$. Using
these assumptions, we can write down the product matrix
\begin{equation}
  \label{eq:3}
  \rho\tilde{\rho} = w_m^2\rho_m\tilde{\rho_m} +
  w_mw_n(\rho_m\tilde{\rho_n}+\rho_n\tilde{\rho_m}) +
  w_n^2\rho_n\tilde{\rho_n}.
\end{equation}
Our condition (\ref{eq:condition}) makes the cross terms drop out,
since
\begin{equation}
  \label{eq:4}
  \rho_i\tilde{\rho_j} =
  |\alpha_i\rangle\langle\alpha_i|\hat{\sigma}_y\otimes\hat{\sigma}_y|\alpha_j\rangle\langle\alpha_j|\hat{\sigma}_y\otimes\hat{\sigma}_y = 0.
\end{equation}
Furthermore, condition (\ref{eq:condition}) together with
Eq. (\ref{eq:1}) gives the following relation
\begin{eqnarray}
  \label{eq:5}
  \rho_i\tilde{\rho_i}|u_{j0}\rangle &&=
  |\alpha_i\rangle\langle\alpha_i|\hat{\sigma}_y\otimes\hat{\sigma}_y|\alpha_i\rangle\langle\alpha_i|\hat{\sigma}_y\otimes\hat{\sigma}_y|\alpha_j\rangle \nonumber
  \\
  &&=
  \delta_{ij}C_j^2|u_{j0}\rangle.
\end{eqnarray}
The Eqs. (\ref{eq:3})-(\ref{eq:5}) give two of the four
eigenequations of the product matrix
\begin{equation}
  \label{eq:6}
  \rho\tilde{\rho}|v_i\rangle =
  (\delta_{im}w_m^2C_m^2+\delta_{in}w_n^2C_n^2)|v_i\rangle,
\end{equation}
where $|v_i\rangle = |u_{i0}\rangle = |\alpha_i\rangle$. Since
these two eigenvectors only span two dimensions in the
four-dimensional space, we can always find another two vectors
$|\alpha_k^\perp\rangle, k = 2,3$ which are linear independent of
each other and orthogonal to the two eigenstates. Thus the last two eigenvectors are $|v_k\rangle =
\hat{\sigma}_y\otimes\hat{\sigma}_y|\alpha_k^\perp\rangle, k =2,3$, both with zero 
eigenvalue. Eq. (\ref{eq:concurrence}) now gives
our mixed concurrence formula.
\\
\\
\emph{Theorem 1} applies to any system with a mixture of two
pure states satisfying condition (\ref{eq:condition}). It can easily be extended to apply to the mixing of more
pure states. The requirement is then that condition
(\ref{eq:condition}) must hold for all pairs of pure states.

In our case, it would have been interesting if, when including
all four levels, the concurrence could be calculated as
\begin{equation}
  \label{eq:weightconcurrence4}
  C = \max_k\{2w_kC_k - \sum_iw_iC_i,0\}.
\end{equation}
In fact, for a three-level approximation (involving the first
three levels), this formula is correct, however the exact four-level
concurrence is not in agreement with
Eq. (\ref{eq:weightconcurrence4}), because condition
(\ref{eq:condition}) does not hold for mixing the ground state with the third excited state.

\section{Arbitrary fields}
\label{theta}
In section \ref{2-qubits} we treated a case of a magnetic field
orthogonal to the Ising direction. We are now going to generalize
this to arbitrary magnetic fields. The new Hamiltonian can be
written
\begin{eqnarray}
  \label{eq:HamiltonianTheta}
  \hat{H} =&& 2J\,\hat{\sigma}_z\otimes\hat{\sigma}_z \nonumber
  \\
  &&+
  B\sin\theta\,(\hat{\sigma}_x\otimes\hat{I}+\hat{I}\otimes\hat{\sigma}_x)
  \nonumber
  \\
  &&+
  B\cos\theta\,(\hat{\sigma}_z\otimes\hat{I}+\hat{I}\otimes\hat{\sigma}_z),
\end{eqnarray}
where $\theta$ is the angle between the magnetic field and the Ising
direction. It is sufficient to consider variation of $B$ in a plane containing
the Ising direction, because in 3 spatial dimensions the Hamiltonian
possesses rotational symmetry about the $z$ axis.

\begin{figure}
    \begin{minipage}{\columnwidth}
    \begin{center}
        \resizebox{1.0\columnwidth}{!}{\includegraphics{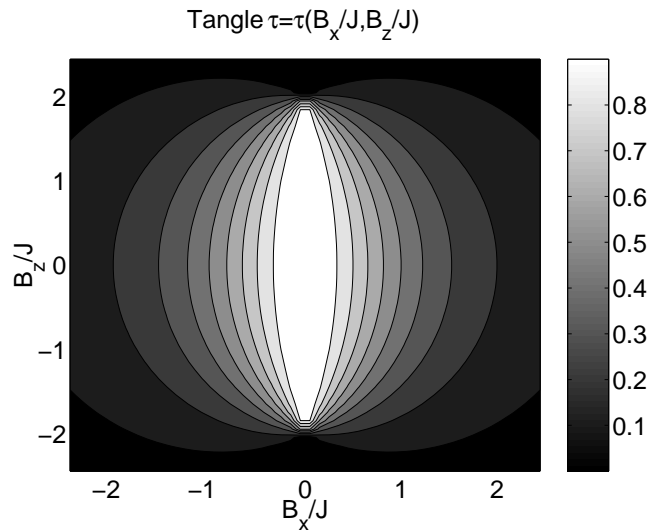}}
    \end{center}
    \end{minipage}
    \caption{Contour plot of the tangle at zero temperature in a Cartesian coordinate
      system. Note that the line of non-entangled states at $B_x = 0$ for $|B_z| < 2J$ can not be seen.}
    \label{fig:3}
\end{figure}
\begin{figure}
    \begin{minipage}{\columnwidth}
    \begin{center}
        \resizebox{1.0\columnwidth}{!}{\includegraphics{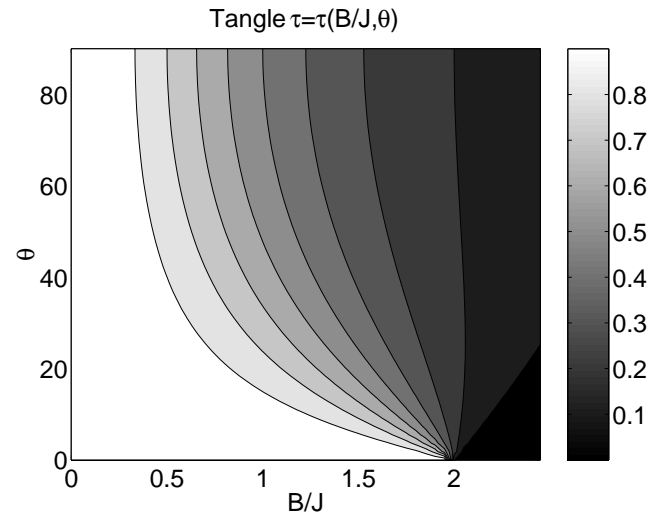}}
    \end{center}
    \end{minipage}
    \caption{Contour plot of the tangle at zero temperature in a
      spherical coordinate system.}
    \label{fig:4}
\end{figure}

The expression for the tangle is analytically solvable. However, because of a
difficult cubic equation in the diagonalization, the expressions are
complicated, so we present the results in graphical form. At zero
temperature, Fig. \ref{fig:3} shows
our solution when the tangle is plotted as a function of
$B_x=B\sin{\theta}$ and $B_z=B\cos{\theta}$, and Fig. \ref{fig:4}
shows the solution when the tangle is plotted as a function of the
amplitude $B$ and angle $\theta$.

We notice that the region around $B_x = 0$ for all $|B_z| < 2J$
has the highest possible entanglement. At exactly at $B_x=0$ there
should not be {\em any} entanglement, the white region of
Fig. \ref{fig:3} indicates a quantum phase transition at $B_x=0$
(the sharpness of the transition being illustrated by the fact
that the zero entanglement line at $B_x=0$ is so thin that it is
invisible). For small angles, $\theta$, there are two energy
levels close to the energy $-2 J$ with corresponding states close
to the Bell states $|\Psi^{\pm}\rangle = (|01\rangle\pm |10\rangle)/\sqrt(2)$. Thus we get a maximally
entangled qubit pair in the limit. However, at $\vec{B} = B_z
\vec{z}$, with $|B_z| < 2J$, the states are degenerate with no
entanglement as a result. In the case $B_z > 2 J$ ($B_z < -2J$)
the ground state is always the non-entangled state $|11\rangle$
($|00\rangle$). Also notice that $\theta = \pi/2$ corresponds to
our earlier orthogonal case, thus the tangle follows
Eq. (\ref{eq:tangle}). Even when $|B_x|$ is increased to the point
where it starts to dominate, the spins will simply align along $B_x$
and give a disentangled state. Thus the entanglement falls off
with increasing strength of the magnetic field in either
direction.

\begin{figure}
    \begin{minipage}{\columnwidth}
    \begin{center}
        \resizebox{1.0\columnwidth}{!}{\includegraphics{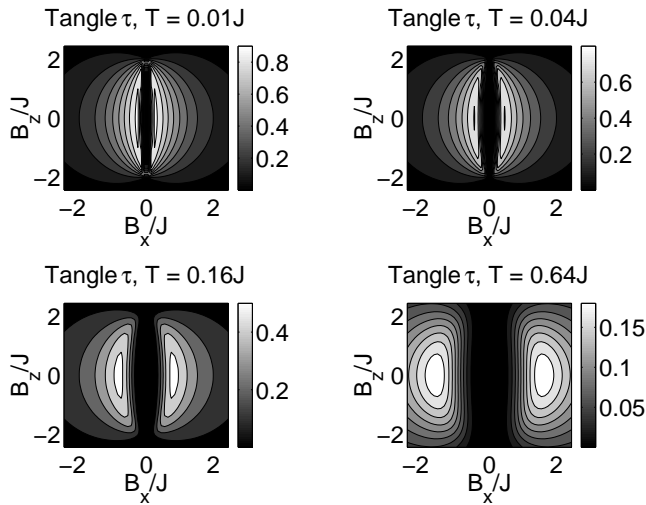}}
    \end{center}
    \end{minipage}
    \caption{Contour plots of the tangle in Cartesian coordinates for
      various finite temperatures, $T$.}
    \label{fig:5}
\end{figure}

\begin{figure}
    \begin{minipage}{\columnwidth}
    \begin{center}
        \resizebox{1.0\columnwidth}{!}{\includegraphics{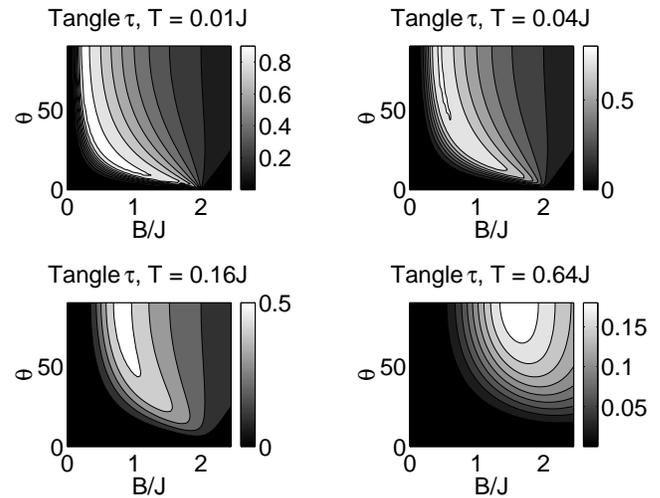}}
    \end{center}
    \end{minipage}
    \caption{Contour plots of the tangle in spherical coordinates. The angle, resulting in the maximum entanglement for a given magnetic field amplitude, varies as a function of the amplitude and temperature.}
    \label{fig:6}
\end{figure}

Let us now look at the case of finite temperature (thermal
entanglement). The first excited state is $|\Psi^{-}\rangle$
which lies at the energy $-2J$. This state is totally independent
of magnetic field, thus the tangle corresponding to this state
forms a constant plane at $1$. Fortunately,
condition (\ref{eq:condition}) in our theorem is also satisfied,
which makes Eq. (\ref{eq:weightconcurrence2}) valid. In
Figs. \ref{fig:5} and \ref{fig:6} a numerical solution is shown at
a low temperature. Note how fast the tangle drops to zero for a
low $|B_x|$-component. This does not contradict the
concurrence mixing formula, as the weights $w_i$ also depend on
the magnetic field through the energy. The fast drop in the tangle is
due to the degeneracy at $B_x = 0$. The smallness of the energy
difference at low values of $|B_x|$ makes the two levels almost
equally populated even for small temperatures. The line of zero
entanglement at $B_x = 0$ for $T=0$ has broadened into a region of
almost zero entanglement in the finite temperature case (compare
Figs. \ref{fig:3}, \ref{fig:5}).

It is also interesting to see that there exists an angle
$\theta = \theta^*(B,T)$, where the entanglement is maximum for a
given temperature and amplitude
\begin{equation}
  \label{eq:maxtangle}
  \max \tau(\vec{B},T) = \tau(\theta^*(B,T)).
\end{equation}
This feature can be explained heuristically if we assume
that with $B_x$ and $B_z$ fixed, the entanglement should
change continuously with temperature. We know that
increasing the temperature widens the low entanglement zone
around $B_x=0$ and the entanglement has to fall off for large
$|B_x|$. So it is expected that at some intermediate value of $|B_x|$
(and hence $\theta$) the maximal entanglement will be reached. As
we increase the temperature further, the near-zero entanglement zone
centered around $B_x=0$ widens even more and pushes the
entanglement maxima away to higher and higher values of $|B_x|$.
The highest value of the tangle tends more and more towards
orthogonal fields (c.f. 
Figs. \ref{fig:5}, \ref{fig:6}). The
preferred angle traverses from $\theta = 0$ at zero temperature
(Fig. \ref{fig:4}) to $\theta = \pi/2$ at $T\approx J$
(Fig. \ref{fig:6}). In the classical limit, i.e. at very high
temperatures all entanglement fades away as expected, because the
state is completely mixed.

We can use the two-level approximation to get an estimate of the angle 
creating maximal entanglement. Let us first
estimate the ground state energy, $\epsilon$. Since we know that
the lowest two energy levels are very close and the first excited state is
$-2J$, we can use the approximation $-2J+\epsilon \approx -4J$
while solving for eigenvalues of the
Hamiltonian (\ref{eq:HamiltonianTheta}). The ground state energy is then
\begin{equation}
  \label{eq:Gndstateenergy}
  \epsilon = -2J - \frac{4B_x^2}{4J^2-B_z^2}J.
\end{equation}
For the above approximation to hold, we must
operate in a region not too close to the poles at $B_z = \pm 2J$. This gives the following energy difference between the two lowest levels
\begin{equation}
  \label{eq:Energydifference}
  \Delta\epsilon = -2J - \epsilon \equiv \frac{4B_x^2}{4J^2-B_z^2}J.
\end{equation}
The two terms of the ground state energy (\ref{eq:Gndstateenergy}) can
be considered as the first two terms in a Taylor expansion. This
suggests that we instead write
\begin{equation}
  \label{eq:Energy}
  \epsilon = -2J\sqrt{1+\frac{\Delta\epsilon}{J}} = -2J\sqrt{1+\frac{4B_x^2}{4J^2-B_z^2}}.
\end{equation}
Let us, from here onwards, measure all energies in units of $J$,
i.e. let $J=1$. For a magnetic field only in the $x$ direction, the
concurrence of the ground state is $C_0 = \tau^{1/2} = -2J/\epsilon$,
where $\tau$ is given by Eq. (\ref{eq:tangle}). For non-zero $B_z$
this formula remains an excellent approximation (as we have verified
numerically). Substituting Eq. (\ref{eq:Energy}) for $\epsilon$ gives
\begin{equation}
  \label{eq:GeneralConcurrence}
  C_0 = \frac{1}{\sqrt{1+\Delta\epsilon}}.
\end{equation}
Recall that the first excited state is $|\Psi^-\rangle$ with
concurrence $C_1 = 1$.
From the concurrence mixing theorem (\ref{eq:theorem}) we get an
approximation of the thermal concurrence. If $w_i$ are weights
following Maxwell-Boltzmann statistics, then the maximum entanglement is
reached when the following condition holds:
\begin{equation}
  \label{eq:MaxCondition}
  T(1+e^{\Delta\epsilon/T}) = 2(1+\Delta\epsilon+\sqrt{1+\Delta\epsilon}).
\end{equation}
In the region we are interested in, where temperature is not
bigger than $\sim 20\%$ of the coupling constant, $J$, the temperature
is also much smaller than $\Delta\epsilon$, and Eq. (\ref{eq:MaxCondition}) can be further simplified to
\begin{equation}
  \label{eq:ApproxCondition}
  \Delta\epsilon = T \ln\frac{4}{T}.
\end{equation}
Remember that the ground state energy is still a function of the
magnetic field.  In order to fix the temperature,
$\Delta\epsilon$ has to be fixed, i.e. maximum entanglement is reached
in the cross section between the energy surface and the constant
energy plane that follows from Eq. (\ref{eq:ApproxCondition}).
From Eq. (\ref{eq:Energydifference}) it is clear that this is described
by an ellipse in a $B_xB_z$-plane. Using the field amplitude $B$ as
a parameter, the angle we defined in Eq. (\ref{eq:maxtangle})
is given by
\begin{equation}
  \label{eq:thetafound}
  \sin\theta^* = \pm\sqrt{\frac{\Delta\epsilon}{4-\Delta\epsilon}\frac{4-B^2}{B^2}},
\end{equation}
assuming that the parameter $|B|>\sqrt{\epsilon}$.

Another way to parameterize the optimum line is to let $B_z$ be the
parameter and solve for $B_x$,
\begin{equation}
  \label{eq:Bxfound}
  B_x = \pm\sqrt{\Delta\epsilon\left(1-\frac{B_z^2}{4}\right)},
\end{equation}
with $|B_z| < 2$. Again keep in mind that Eqs. (\ref{eq:thetafound})
and (\ref{eq:Bxfound}) follow from the
assumption that $\Delta\epsilon\approx 0$, and therefore the
ellipse is not closed around the poles at $B_z = \pm 2$
(\emph{cf} Figs. \ref{fig:5}, \ref{fig:7}).

\section{Qubit rings}
\label{many}

\begin{figure}
    \begin{minipage}{\columnwidth}
    \begin{center}
        \resizebox{1.0\columnwidth}{!}{\includegraphics{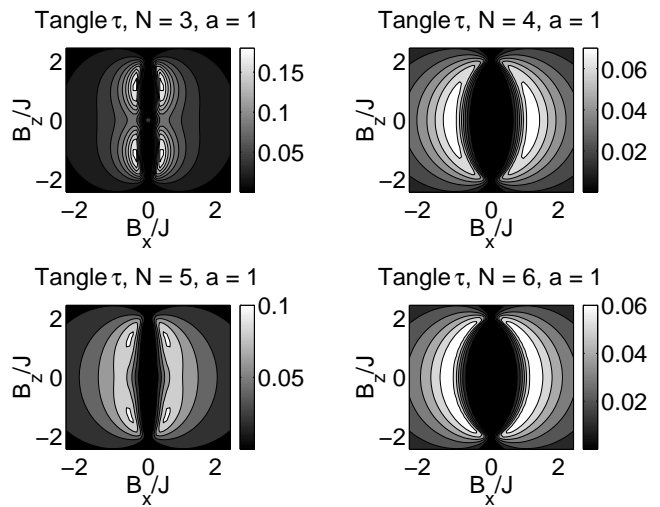}}
    \end{center}
    \end{minipage}
    \caption{Contour plots of the tangle at temperature $T = 0.10J$. N is the number of
      qubits in the chain and a=1 means that entanglement is measured
      between two neighboring qubits.}
    \label{fig:7}
\end{figure}

As the expressions for the thermal tangle of the general two-qubit
case are already quite complicated, we can not expect to find any
easily manageable analytic expressions in the many-qubit case.
Instead, we have performed numerical simulations, which gives the
entanglement between neighboring qubits as shown in
Fig. \ref{fig:7}.

The behavior for even $N$ rings is quite similar to
that of the two qubit case. To understand why there is an extra
low entanglement zone around $B_z=0$ in the case of odd $N$
rings, one has to go back the basic cause for entanglement
arising in the Ising chain. It results from the
competition between the term $\hat{\sigma}_z\otimes\hat{\sigma}_z
+ B_z \hat{\sigma}_z$ trying to impose spin order in the $z$
direction and $B_x \hat{\sigma}_x$ trying to impose spin order in
the $x$ direction. In the odd qubit case, it is impossible for
all neighboring spins to be oriented oppositely, so the ordering
power of $\hat{\sigma}_z\otimes\hat{\sigma}_z$ is significantly
reduced. In such circumstances, it is mainly the competition
between $B_z \hat{\sigma}_z$ and $B_x \hat{\sigma}_x$ (albeit
aided by the small $\hat{\sigma}_z\otimes\hat{\sigma}_z$
interaction) which determines the entanglement. Thus the high
entanglement values near $B_z=0$ present for the two qubit (and
all other even $N$) cases disappear for odd $N$. Note
also that the entanglement in the odd $N$ case is somewhat larger
in magnitude compared with the even $N$ case. This is a result of
the fact that the two terms $\hat{\sigma}_z\otimes\hat{\sigma}_z$
and $B_z \hat{\sigma}_z$ compete for the type of $z$ ordering
(parallel or anti-parallel neighboring spins). In even $N$ case,
this competition is much stronger and this tends to lower the
entanglement by reducing the net effect of $z$ ordering terms
with respect to $x$ ordering terms. As the number $N$ of qubits
in the chain is increased, the difference between even and odd
$N$ chains should disappear (because for large $N$, adding or
removing an extra qubit from the chain should not make a
significant difference). This effect is clearly seen in
Fig. \ref{fig:7} where the difference in appearance
between the plots for $N=3$ and $N=4$ is much greater than that
between $N=5$ and $N=6$.

\begin{figure}
    \begin{minipage}{\columnwidth}
    \begin{center}
        \resizebox{1.0\columnwidth}{!}{\includegraphics{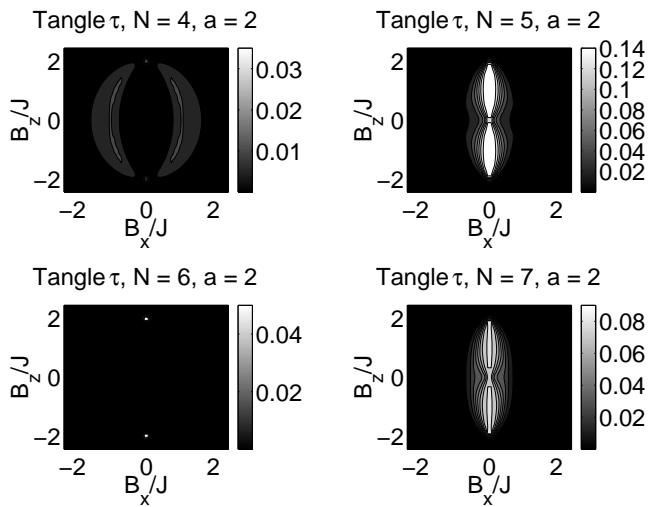}}
    \end{center}
    \end{minipage}
    \caption{Contour plots of the tangle at temperature $T = 0.10J$. N is the number of
      qubits in the chain and a=2 means that entanglement is measured
      in pairs with one qubit in-between.}
    \label{fig:8}
\end{figure}
Entanglement can also be calculated between non-neighboring qubits
with the results shown in Fig. \ref{fig:8} for next-nearest neighbors.
Again we observe that the even $N$ case has lower entanglement on
average than the odd $N$ case. Also in the odd qubit case, the
entanglement between next-nearest neighbors is somewhat
complementary to that between nearest neighbors (this can be seen for
example by placing the plots for $N=5$ in the two cases on top of each
other). Thus the amount of entanglement between pairs of nearest
neighbors and pairs of next-nearest neighbors can be controlled by
varying the field direction.

\section{Conclusions}

In this paper we have investigated the natural thermal
entanglement arising in an Ising ring with a magnetic field in
an arbitrary direction. We have investigated two qubit
analytically and three through seven qubits
numerically. One of the most interesting results is the fact that for
a given temperature, the (nearest neighbor, pairwise) entanglement in the ring can be
maximized by rotating the magnetic field (at fixed magnitude)
to an optimal direction. This can be regarded as {\em
magnetically induced} entanglement. The pairwise entanglement between
next-nearest neighbors can be maximized by rotating the field to a
different direction. We have also proved a theorem of
mixing of concurrences which is applicable to any system in which the
pure states in the mixture have no spin-flip overlap.

So far we have have only considered pairwise entanglement. In future
work we will estimate the entanglement between three or more 
qubits in the ring and also focus on investigating ways to detect the natural
entanglement in Ising models, and on investigation of the
entanglement in the large variety of available condensed matter models
of interacting systems.

\begin{acknowledgments}
This work was funded by the UK Engineering and Physical Sciences
Research Council and the European Union Project EQUIP (contract IST-1999-11053).
\end{acknowledgments}


\end{document}